\documentclass[twocolumn]{webofc}
\usepackage[varg]{txfonts}   
\usepackage{bm}
\usepackage{color}
\newcommand{\pprime}{{\prime\prime}}

\begin{document}
\title{Rheology of dilute granular gases with hard-core and inverse power-law potentials}
\author{
    \firstname{Yuria} \lastname{Kobayashi}\inst{1} 
    \fnsep
    \thanks{\email{yuria@st.go.tuat.ac.jp}}
    \and
    \firstname{Shunsuke} \lastname{Iizuka}\inst{1} \and
    \firstname{Satoshi} \lastname{Takada}\inst{1}\fnsep
    \thanks{\email{takada@go.tuat.ac.jp}, corresponding author}
}

\institute{
    Department of Mechanical Systems Engineering, 
    Tokyo University of Agriculture and Technology, 
    2-24-16 Naka-cho, Koganei, Tokyo 184-8588, Japan
}

\abstract{
The kinetic theory of dilute granular gases with hard-core and inverse power-law potentials is developed.
The scattering process is studied theoretically, which yields the relative speed and the impact parameter dependence of the scattering angle.
The viscosity is derived from the Boltzmann equation and its temperature dependence is plotted.
We also perform the direct simulation Monte Carlo to check the validity of the theory.
}
\maketitle


\section{Introduction}
Understanding the flow behavior of granular gases is important not only from a physical perspective but also due to its industrial significance. 
While extensive theoretical studies have been conducted on the rheology of granular gases treated as rigid spheres, new physics emerges when particles acquire electric charges through processes like triboelectric charging. 
Developing theories to address such phenomena is an urgent challenge.

In general, for the homogeneous cooling process of hard-core sphere systems, it is well known that Haff's law applies, where the temperature $T$ decreases over time as $T\sim t^{-2}$ in the long-time limit~\cite{Haff83}. 
However, in the case of charged particles, dissipation ceases when the kinetic energy becomes smaller than the inter-particle potential energy, as particles are unable to approach each other's cores.

Several studies have explored related phenomena~\cite{Scheffler02, Poschel03, Takada17, Takada22, Yoshii23, Takada24, Singh18, Singh19}. 
For instance, theoretical analyses have been conducted using models that mimic charged particles by varying the restitution coefficient based on collision velocity, even when treating the particles as hard-core spheres~\cite{Scheffler02, Poschel03, Takada17}. 
These studies successfully derived time-evolution equations for temperature, which showed excellent agreement with results obtained from Direct Simulation Monte Carlo (DSMC) methods~\cite{Takada17, Takada22}. 
Another line of research has introduced models incorporating scattering processes of charged particles by considering square-well and square-shoulder potentials~\cite{Yoshii23}. 
These models also demonstrated strong agreement with molecular dynamics (MD) simulations.

In a recent study~\cite{Takada24}, analyses are conducted on systems where particles interact through repulsive inverse power-law potentials. 
While the study discussed aspects such as linear stability analysis, it employed a model in which particles undergo inelastic collisions despite the absence of hard cores. 
This approach neglects the deceleration effects naturally caused by inter-particle repulsion, rendering the model somewhat unphysical.

In the present work, we propose a more realistic model by incorporating hard-core particles into the framework of~\cite{Takada24}. 
Specifically, we aim to analyze dilute systems under uniform shear flow using the Boltzmann equation to derive key transport coefficients, such as viscosity, and provide a more accurate representation of charged granular gases.
We also perform the direct simulation Monte Carlo to check the validity of the theory.

\section{Model and setup}
\begin{figure}[htbp]
    \centering
    \includegraphics[width=\linewidth]{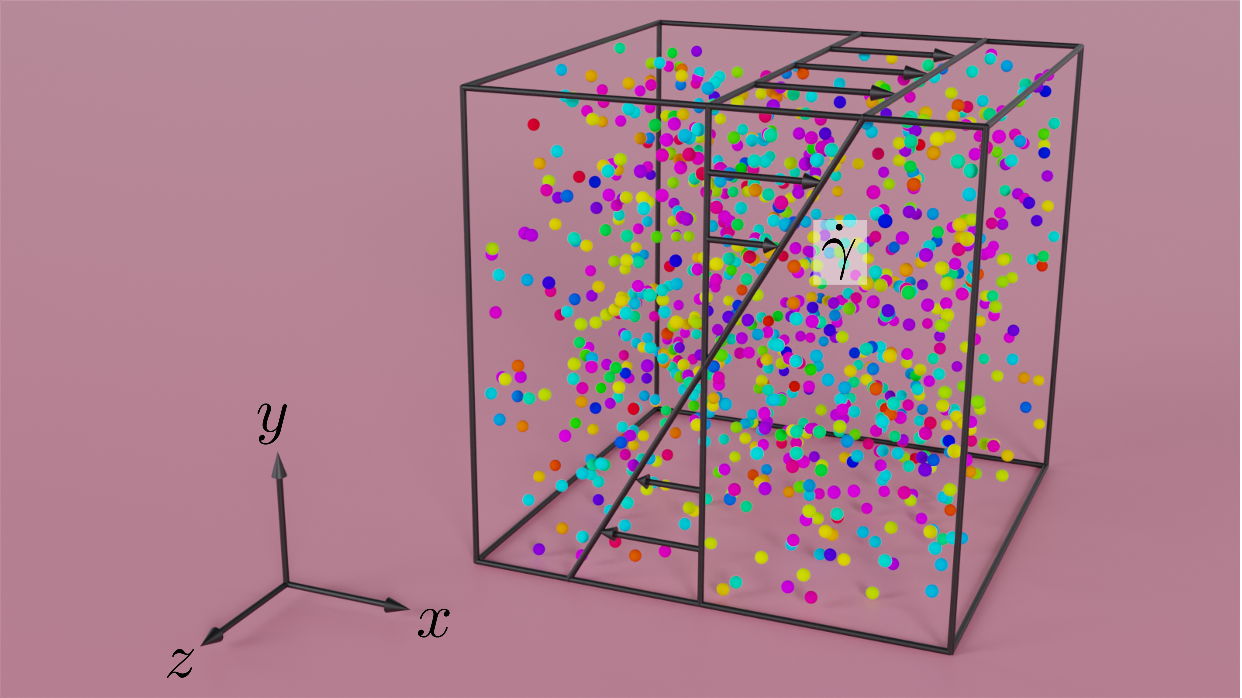}
    \caption{Schematic picture of the system.
    A shear is applied in the $x$-direction with the shear rate $\dot\gamma$.
    }
    \label{fig:system}
\end{figure}

\begin{figure}[htbp]
    \centering
    \includegraphics[width=0.7\linewidth]{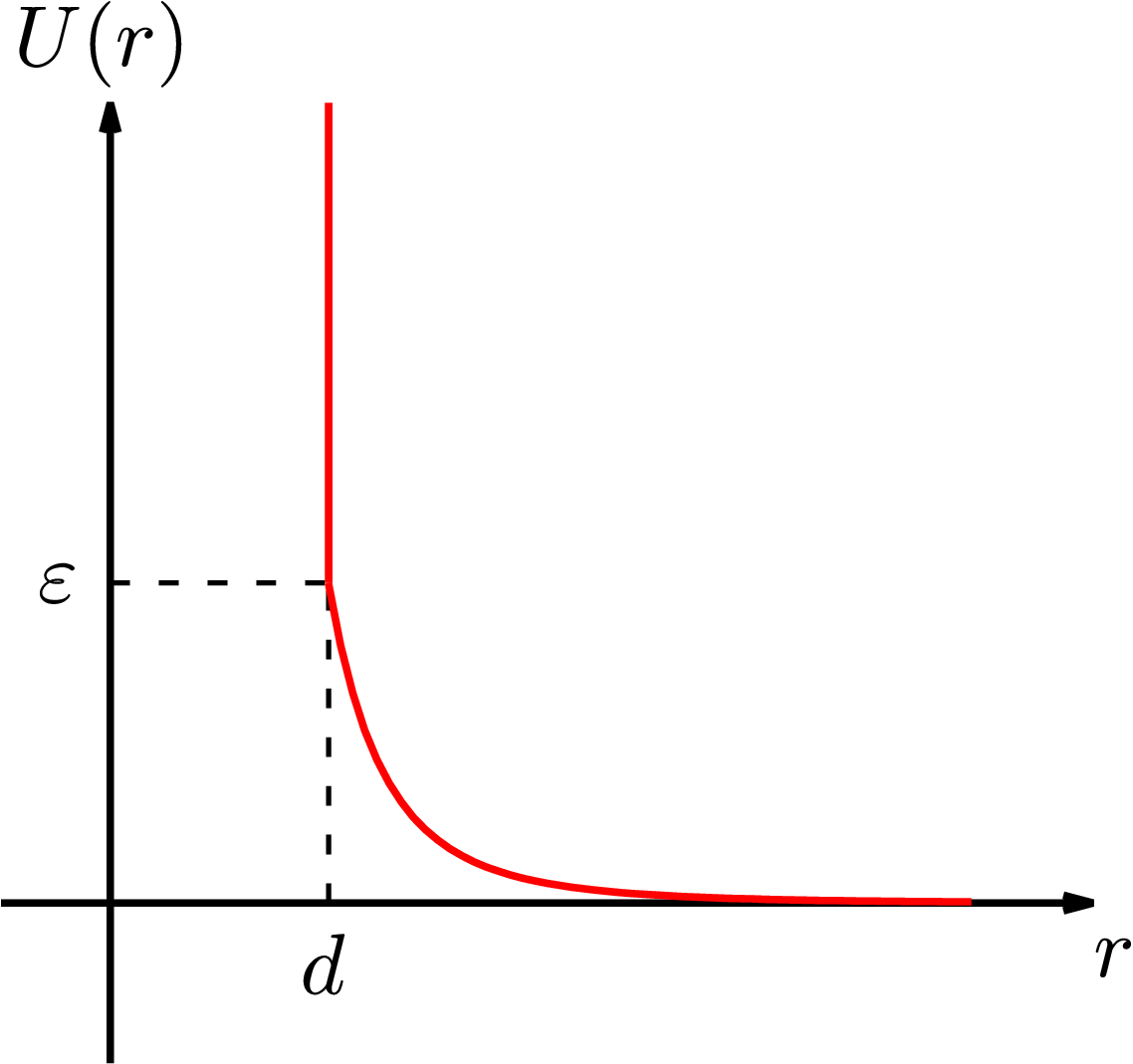}
    \caption{Schematic of the potential.
    Hard-core potential with the inverse power-law part for $r>d$ (red solid line) given by Eq.~\eqref{eq:potential}.
    }
    \label{fig:potential}
\end{figure}

Let us consider the system where monodisperse particles whose mass and diameter are given by $m$ and $d$, respectively, distribute uniformly with the number density $n$ as shown in Fig.~\ref{fig:system}.
Here, we assume that the number density is small enough ($nd^3\ll 1$).
The interparticle potential between particles is the sum of the hard-core and the inverse power-law repulsive part as
\begin{equation}
    U(r) = 
    \begin{cases}
        \infty & (r\le d)\\
        \varepsilon\left(\dfrac{d}{r}\right)^\alpha & (r>d)
    \end{cases},
    \label{eq:potential}
\end{equation}
(see Fig.~\ref{fig:potential}).
Now, $\alpha> 2$ is assumed because we want the condition $4\pi r^2 U(r)\to0$ at infinity ($r\to\infty$).
When the relative distance between two particles reaches $r=d$, the particles have an inelastic collision and repel with a restitution coefficient $e_0<1$.

\section{Scattering process}
First, let us consider the scattering process~\cite{Goldstein} when two particles interact with each other.
When the relative speed and the impact parameter are given by $\bm{v}$ and $b$, respectively, the angle $\theta$ between the
incidental asymptote and the closest approach $\theta_\alpha$ is given by
\begin{align}
    \theta_\alpha
    &= b \int_0^{1/r_\mathrm{min}}
    \frac{\mathrm{d}u}{\sqrt{1-b^2u^2-\dfrac{4}{mv^2}U\left(\dfrac{1}{u}\right)}}\nonumber\\
    &= \int_0^{x_0}
    \frac{\mathrm{d}x}{\sqrt{1-x^2-\left(\dfrac{x}{\tilde{b}}\right)^\alpha}},
    \label{eq:theta}
\end{align}
where $r_\mathrm{min}\ge d$ is the closest distance, $x_0=b/r_\mathrm{min}$, and 
\begin{equation}
    \tilde{b}\equiv
    \left(\frac{mv^2}{4\varepsilon}\right)^{1/\alpha}\frac{b}{d}.
\end{equation} 
We note that the integral~\eqref{eq:theta} can be explicitly evaluated for certain values of $\alpha$~\cite{Goldstein, Takada24}.

Depending on $b$ and $v$, two types of scattering occur.
These are (i) when the two particles reach each other's hard-cores and inelastic collisions occur, and (ii) when they are scattered by repulsive force before they reach the hard-cores.
The former inelastic collision occurs when
\begin{equation}
    v\ge \sqrt{\dfrac{4\varepsilon}{m}},\quad
    b\le \nu_\mathrm{r}d,
\end{equation}
are satisfied, where $\nu_\mathrm{r}$ is the refractive index given by
\begin{equation}
    \nu_\mathrm{r}
    \equiv
    \sqrt{1-\dfrac{4\varepsilon}{mv^2}}.
\end{equation}
We can also determine $x_0$ as
\begin{equation}
    x_0 = 
    \begin{cases}
        \dfrac{b}{d} & \left(v\ge \sqrt{\dfrac{4\varepsilon}{m}},\ b\le \nu_\mathrm{r}d\right)\\
        x_\mathrm{el} & (\textrm{otherwise})
    \end{cases},
\end{equation}
where $x_\mathrm{el}$ is the solution of the equation in the square root of the denominator of Eq.~\eqref{eq:theta}.

\section{Boltzmann equation}
Let us consider the situation where the velocity distribution function $f(\bm{v},t)$ satisfies the Boltzmann equation
\begin{equation}
    \left(\frac{\partial}{\partial t}
    +\bm{v}_1\cdot \bm\nabla\right)f(\bm{v}_1,t)
    =J(f,f),
    \label{eq:Boltzmann_eq}
\end{equation}
where $J(f,f)$ is the collision integral given by
\begin{align}
    J(f,f)
    &=\int \mathrm{d}\bm{v}_2 \int \mathrm{d}\hat{\bm{k}} \sigma(\chi,\bm{v}_{12}) v_{12}\nonumber\\
    &\hspace{1em}\times\left[\frac{1}{\mathcal{E}^2} f(\bm{v}_1^\pprime,t)f(\bm{v}_2^\pprime,t) 
    - f(\bm{v}_1,t)f(\bm{v}_2,t)\right].
\end{align}
Here, the relationship between the pre-collisional velocities $\bm{v}_1^\pprime, \bm{v}_2^\pprime$ and the post-collisional velocities $\bm{v}_1, \bm{v}_2$ is written as
\begin{equation}
    \displaystyle \bm{v}_1 
    = \bm{v}_1^\pprime 
    - \dfrac{1+\mathcal{E}}{2}\left(\bm{v}_{12}^\pprime \cdot \hat{\bm{k}}\right)\hat{\bm{k}},\quad
    \displaystyle \bm{v}_2 
    = \bm{v}_2^\pprime 
    + \dfrac{1+\mathcal{E}}{2}\left(\bm{v}_{12}^\pprime \cdot \hat{\bm{k}}\right)\hat{\bm{k}},
    \label{eq:pre_post}
\end{equation}
with 
\begin{equation}
    \mathcal{E} =
    \begin{cases}
        \sqrt{1-(1-e_0^2)\nu_\mathrm{r}^2} & \left(v\ge \sqrt{\dfrac{4\varepsilon}{m}},\ b\le \nu_\mathrm{r}d\right)\\
        1 & (\textrm{otherwise})
    \end{cases}.
    \label{eq:E}
\end{equation}
We note that $\mathcal{E}$ corresponds to the macroscopic restitution coefficient when even the deceleration and acceleration due to the repulsive potential are taken.

\begin{figure}[htbp]
    \centering
    \includegraphics[width=\linewidth]{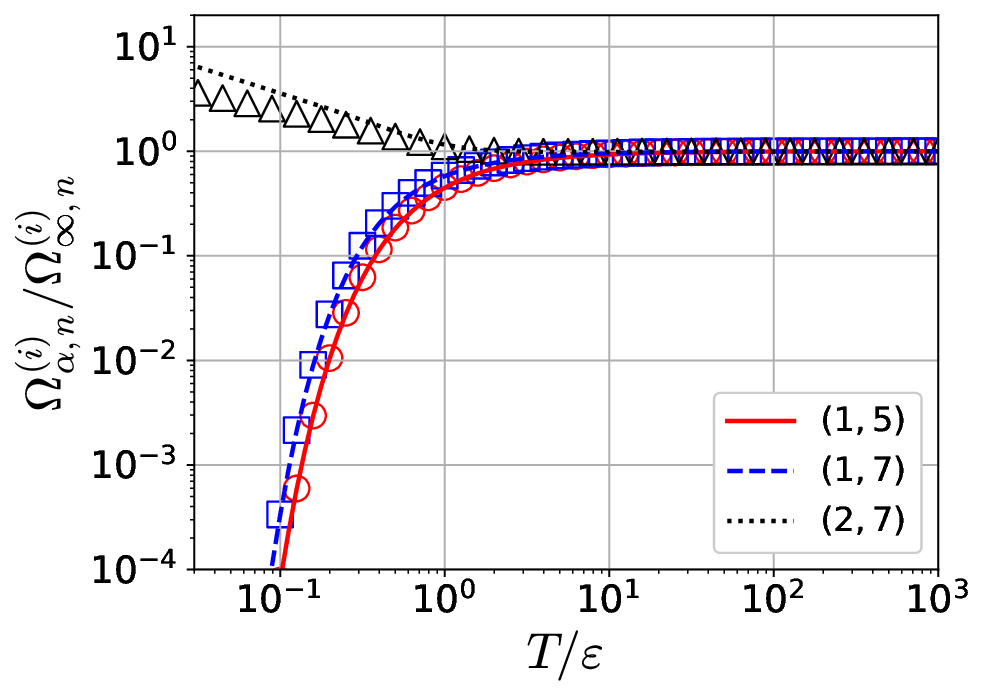}
    \caption{Temperature dependence of $\Omega^{(i)}_{\alpha,n}$ for $\alpha=4$ (line) and $6$ (marker) with $(i,n)=(1,5)$ (red), $(2,5)$ (blue), and $(2,7)$ (black), where $\Omega^{(i)}_{\infty, n}$ is the hard-core limit of $\Omega^{(i)}_{\alpha,n}$.
    Here, the restitution coefficient for inelastic collisions is chosen as $e_0=0.9$.}
    \label{fig:Omega}
\end{figure}

Because we are interested in the uniformly sheared state, let us consider the peculiar velocity $\bm{V}\equiv \bm{v}-\dot\gamma y \hat{\bm{e}}_x$.
Multiplying Eq.~\eqref{eq:Boltzmann_eq} with $mV_\alpha V_\beta$ and integrating over $\bm{V}$, we can obtain the time evolution of the stress as
\begin{equation}
    \partial_t P_{\alpha\beta}
    + \dot\gamma (\delta_{\alpha x}P_{y\beta} + \delta_{\beta x}P_{\alpha y})
    = -\Lambda_{\alpha\beta},
    \label{eq:evol_Pab}
\end{equation}
where
\begin{equation}
    P_{\alpha\beta}
    \equiv m \int \mathrm{d}\bm{V} V_\alpha V_\beta f(\bm{V},t),
\end{equation}
is the $\alpha\beta$ component of the stress and 
\begin{equation}
    \Lambda_{\alpha\beta}
    = -\int \mathrm{d}v_1 m V_\alpha V_\beta
    J(f,f),
\end{equation}
is the moment of the collision integral.
Unfortunately, we do not know the exact solution of the Boltzmann equation.
However, Grad's expansion
\begin{equation}
    f(\bm{v},t)
    =f_\mathrm{M}(\bm{V})\left[1+\frac{m}{2T}\left(\frac{P_{\alpha\beta}}{nT}-\delta_{\alpha\beta}\right)V^2\right],\label{eq:Grad}
\end{equation}
with the Maxwellian
\begin{equation}
    f_\mathrm{M}(V)
    =n\left(\frac{m}{2\pi T}\right)^{3/2}
    \exp\left(-\frac{mV^2}{2T}\right),
\end{equation}
is a good approximation for many systems.
Once we adopt Grad's expansion, we can write the explicit form of $\Lambda_{\alpha\beta}$ as
\begin{equation}
    \Lambda_{\alpha\beta}
    = \zeta p \delta_{\alpha\beta}
    + \nu (P_{\alpha\beta}-p\delta_{\alpha\beta}),
    \label{eq:Lambda_Grad}
\end{equation}
with
\begin{equation}
    \zeta 
    \equiv \frac{\sqrt{2\pi}}{3}
    nd^2 v_\mathrm{T}\Omega_{\alpha,5}^{(1)},\
    \nu
    \equiv \frac{\sqrt{2\pi}}{15}
    nd^2 v_\mathrm{T}
    \left(\Omega_{\alpha,7}^{(1)}
    +\frac{3}{2}\Omega_{\alpha,7}^{(2)}\right),
\end{equation}
where we have introduced 
\begin{subequations}
\begin{align}
    \Omega_{\alpha,n}^{(1)}
    &\equiv \int_0^\infty \mathrm{d}g 
    \int_0^\infty \mathrm{d}b^*
    \left(1-\mathcal{E}^2\right)
    b^* g^n \cos^2\theta_\alpha
    \mathrm{e}^{-g^2/2},\\
    \Omega_{\alpha,n}^{(2)}
    &\equiv \int_0^\infty \mathrm{d}g 
    \int_0^\infty \mathrm{d}b^*
    \left(1+\mathcal{E}\right)^2
    b^* g^n \sin^2\theta_\alpha \cos^2\theta_\alpha
    \mathrm{e}^{-g^2/2}.
\end{align}
\end{subequations}
The temperature dependencies of these quantities are plotted in Fig.~\ref{fig:Omega} with $\Omega_{\infty, 5}^{(1)}= 2(1-e_0^2)$, $\Omega_{\infty, 7}^{(1)}= 12(1-e_0^2)$, and $\Omega_{\infty, 7}^{(2)}= 4(1+e_0)^2$.
In the hard-core limit, $\zeta$ and $\nu$ becomes $\zeta \to (4/3) nd^2 \sqrt{\pi T/m}(1-e_0^2)$ and $\nu\to (4/5)nd^2\sqrt{\pi T/m}(1+e_0)(3-e_0)$, respectively.
As shown in Fig.~\ref{fig:Omega}, $\Omega_{\alpha,n}^{(1)}$ is insensitive to $\alpha$ while $\Omega_{\alpha,n}^{(2)}$ depends on $\alpha$ in the low-temperature regime.

Using Eq.~\eqref{eq:Lambda_Grad}, Eq.~\eqref{eq:evol_Pab} becomes
\begin{subequations}\label{eq:dynamic_eqs}
\begin{align}
    &\partial_t T
    + \frac{2}{3n}\dot\gamma P_{xy}
    = -\zeta T,\quad
    \partial_t \Delta T
    + \frac{2}{n}\dot\gamma P_{xy}
    = -\nu \Delta T,\\
    &\partial_t P_{xy}
    + \dot\gamma n\left(T -\frac{1}{3}\Delta T\right)
    = -\nu P_{xy},
\end{align}
\end{subequations}
where we have introduced the anisotropic temperature $\Delta T$ as
\begin{equation}
    \Delta T
    \equiv \frac{P_{xx}-P_{yy}}{n}.
\end{equation}
It should be noted that another anisotropic temperature $\delta T\equiv (P_{xx}-P_{zz})/n$ is important in denser situations~\cite{Hayakawa17, Takada20} while it becomes equivalent to $\Delta T$ in the dilute limit.

\begin{figure}[htbp]
    \centering
    \includegraphics[width=\linewidth]{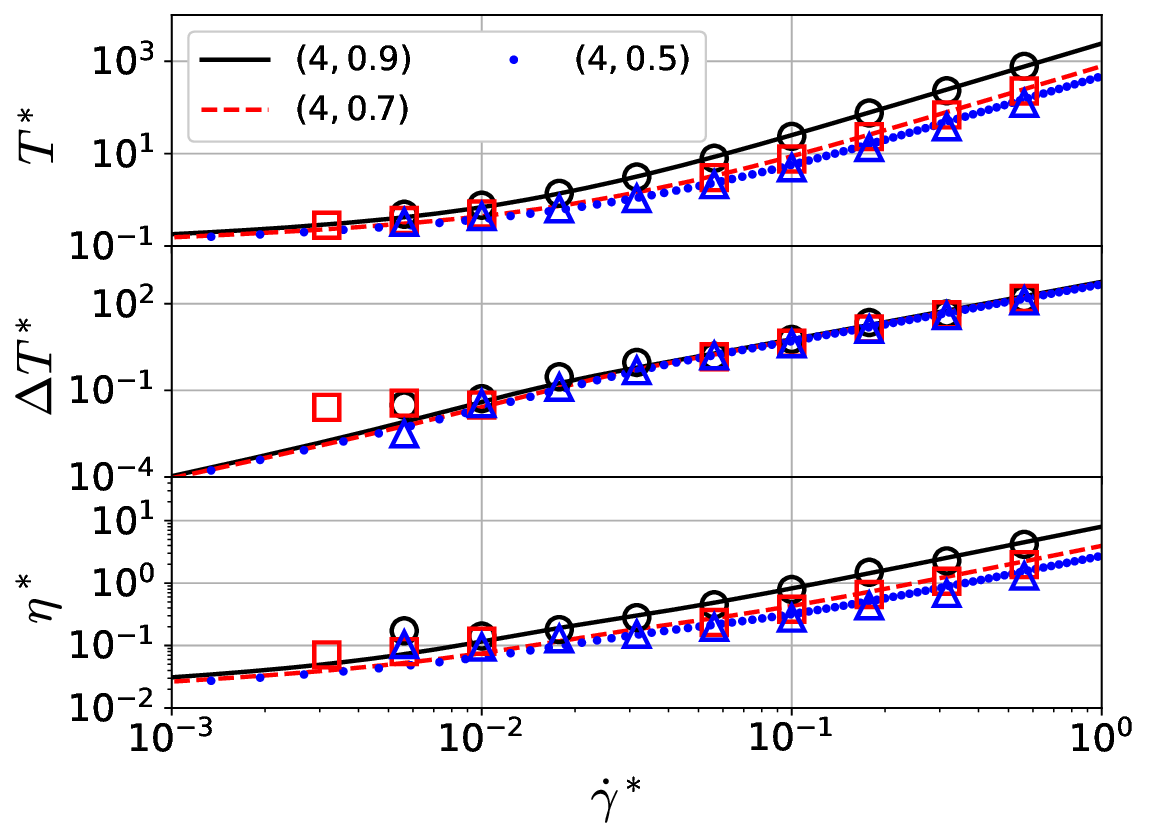}
    \caption{Plots of (top) the temperature $T$, (middle) the anisotropic temperature $\Delta T$, and (bottom) the viscosity $\eta$ against the shear rate $\dot\gamma$ for $(\alpha, e_0)=(4, 0.90)$ (solid lines and open circles), $(4, 0.70)$ (dashed lines and open squares), and $(4, 0.50)$ (dotted lines and open triangles), where we have introduced $T^*\equiv T/\varepsilon$, $\Delta T^*\equiv \Delta T/\varepsilon$, $\eta^*\equiv \eta/(\varepsilon/d^3)$, and $\dot\gamma^*\equiv \dot\gamma (md^2/\varepsilon)^{1/2}$.}
    \label{fig:rheology}
\end{figure}

In this paper, we focus only on the steady-state.
Then, a set of Eqs.~\eqref{eq:dynamic_eqs} yields
\begin{subequations}
\begin{align}
    &\dot\gamma
    = \nu \sqrt{\frac{3}{2}\frac{\zeta}{\nu-\zeta}},\quad
    P_{xy}
    = -\frac{nT}{\nu}
    \sqrt{\frac{3}{2}\zeta(\nu-\zeta)},\\
    &\Delta T 
    = \frac{3\zeta}{\nu}T,\quad
    \eta \equiv -\frac{P_{xy}}{\dot\gamma}
    = nT\frac{\nu-\zeta}{\nu^2}.
\end{align}
\end{subequations}

Figure \ref{fig:rheology} shows the shear rate dependencies of the temperature, anisotropic temperature, and viscosity for $\alpha=4$, where the dimensionless temperature, anisotropic temperature, viscosity, and shear rate are given by $T^*\equiv T/\varepsilon$, $\Delta T^*\equiv \Delta T/\varepsilon$, $\eta^*\equiv \eta/(\varepsilon/d^3)$, and $\dot\gamma^*\equiv \dot\gamma (md^2/\varepsilon)^{1/2}$, respectively.
These quantities are consistent with those of the hard-core system
\begin{subequations}
\begin{align}
    T^{(\mathrm{HC})}
    &= \frac{5(2+e_0)}{12\pi(1-e_0)(1+e_0)^2(3-e_0)^2}
    \frac{m}{n^2 d^4}\dot\gamma^2,\\
    \Delta T^{(\mathrm{HC})}
    &= \frac{25(2+e_0)}{12\pi(1+e_0)^2(3-e_0)^3}
    \frac{m}{n^2 d^4}\dot\gamma^2
    ,\\
    \eta^{(\mathrm{HC})}
    &= \frac{5\sqrt{15}(2+e_0)^{3/2}}{36\pi (1-e_0)^{1/2} (1+e_0)^2(3-e_0)^3}
    \frac{m}{nd^4}\dot\gamma,
\end{align}
\end{subequations}
in the high-shear limit because the repulsive potential ($r>d$) does not play any role in this regime.

\section{Direct simulation Monte Carlo}
To validate the theory explained above, let us perform the direct simulation Monte Carlo (DSMC)~\cite{Bird, Poschel}.
The procedure of the DSMC almost follows that in Ref.~\cite{Poschel}.
Here, the collision rule is the same as Eq.~\eqref{eq:pre_post} with Eq.~\eqref{eq:E}.
In our simulation, the collision parameter $b$ and the collision angle $\theta$ are chosen randomly in the regions $0\le b\le b_\mathrm{max}$ and $0\le \theta < \pi$.
Here, we choose $b_\mathrm{max}=10d$, which is sufficiently large.

We prepare $N=10^3$ monodisperse particles whose mass and diameter $m$ and $d$, respectively, in the cubic box whose linear size is $L$.
The value of $L$ is chosen to satisfy $N/(L/d)^3=0.01$.
A typical snapshot is shown in Fig.~\ref{fig:system}.

The obtained data is plotted in Fig.~\ref{fig:rheology}.
It is obvious that the simulation results show good agreement with the theoretical ones, although there exist some discrepancies in the low-shear regime.
Thus, we conclude that our theory is valid in the wide range of the shear rate.

\section{Conclusion}
In this paper, we have investigated the rheology of dilute granular gases with both hard-core and inverse power-law potentials under sheared flows.
Within the framework of kinetic theory, we have derived the shear rate dependencies of viscosity and temperature.
Our findings indicate that in the high-shear regime, these dependencies remain consistent with those of hard-core gases, whereas deviations become more pronounced in the low-shear regime.
Furthermore, our theoretical predictions have been validated over a wide range of shear rates through direct simulation Monte Carlo.

As a natural extension of our study, the next step is to refine our analysis by incorporating denser systems, which are more representative of realistic physical scenarios.

\section*{Acknowledgment}
This work is partially supported by the Grant-in-Aid of MEXT for Scientific Research (Grant No.~24K06974 and No.~24K07193).



\end{document}